\documentclass{iopart}

\usepackage{graphicx}
\usepackage{array, hhline}
\usepackage{colortbl}
\newcommand{\G}{\cellcolor[gray]{0.7}}
\newcommand{\lG}{\cellcolor[gray]{0.9}}

\newcolumntype{M}[1]{>{\centering\hspace{0pt}\arraybackslash}m{#1}}

\newlength{\mywdithgraph}
\begin{document}

\title [Realization of an Inductance Scale Using a Sampling System] {Realization of an Inductance Scale Traceable to the Quantum Hall Effect Using an Automated Synchronous Sampling System}

\author{F.~Overney and B.~Jeanneret}
\address {Federal Office of Metrology METAS, CH-3003 Bern-Wabern, Switzerland. }
\ead{frederic.overney@metas.ch, blaise.jeanneret@metas.ch}




\begin{abstract}

In this paper, the realization of an inductance scale from 1~$\mu$H to 10~H for frequencies ranging between 50~Hz to 20~kHz is presented. The scale is realized directly from a series of resistance standards using a fully automated synchronous sampling system. A careful systematic characterization of the system shows that the lowest uncertainties, around 12~$\mu$H/H, are obtained for inductances in the range from 10~mH to 100~mH at frequencies in the kHz range. This new measurement system which was successfully evaluated during an international comparison, provides a primary realization of the henry, directly traceable to the quantum Hall effect. An additional key feature of this system is its versatility. In addition to resistance-inductance (R-L) comparison, any kind of impedances can be compared: R-R, R-C, L-L or C-C, giving this sampling system a great potential of use in many laboratories around the world.

\end{abstract}
%
\section{Introduction}
Until the late fifties, the unit of inductance, the henry, was realized by an inductance standard whose impedance was calculated from its geometrical dimensions \cite{Campbell_07}. The farad and the ohm were then obtained from the henry using a series of appropriate bridges. This situation has  radically changed with the development of the Thompson-Lampard calculable capacitor \cite{Thompson56}. The accuracy achieved with the calculable capacitor improved considerably the realization of the farad. As a consequence, the Thomson-Lampard capacitor is still actually used to realize the SI farad from which the SI ohm is deduced. 

Since the nineties, with the advent of the electrical quantum standards, a new representation of the volt and the ohm is based on conventional values of the Josephson and the von Klitzing constants \cite{Taylor89c}. The farad and the henry can then be deduced from the conventional ohm. Today, this approach is used in most of the national metrology institutes for the practical realization of the impedance scales (resistance, capacitance, inductance). 

Although the brides used to link the inductance to the capacitance (traditionally a Maxwell-Wien bridge \cite{Hague71, Zapf61}, a resonance bridge \cite{Rayner80} or a double balance L-bridge  \cite{Callegaro09b}) have the lowest possible relative uncertainties, their manipulation requires special skills and long sequences of manual adjustment. In addition, they are rather dedicated to a unique couple of inductor and capacitor which make the building of the scales rather time consuming. More recently, other versatile methods to realize the inductances scale have been developed like the three-voltmeter method \cite{Callegaro01b} or a digital generator assisted bridge \cite{Corney03}.

In this paper, we propose to realize a representation of the henry directly from the quantum Hall effect using a bridge based on a fully automatic sampling system. Such sampling systems are extremely polyvalent and have been successfully used for the
calibration of power meters \cite{Ramm99}, for the comparison of capacitance to resistance standards \cite{Ramm05} and, more recently, in the development of a Josephson based sinewave synthesizer \cite{Jeanneret09b}. 

In the following sections, the experimental set-up and the measurements procedures are presented together with a detailed discussion of the uncertainty budget. The sampling system has successfully been assessed by an international comparison (EURAMET.EM-S26) where a 100~mH inductance standard was measured at 1 kHz. This comparison was also a good opportunity to extend the frequency range of the inductance measurements between 50~Hz and 20~kHz. In addition, this new system provides a primary realization of the henry, traceable to the quantum Hall effect. 

\section{Description of the sampling system}
\begin{figure*}[tb]
    \begin{center}
        \includegraphics[width=0.7\textwidth] {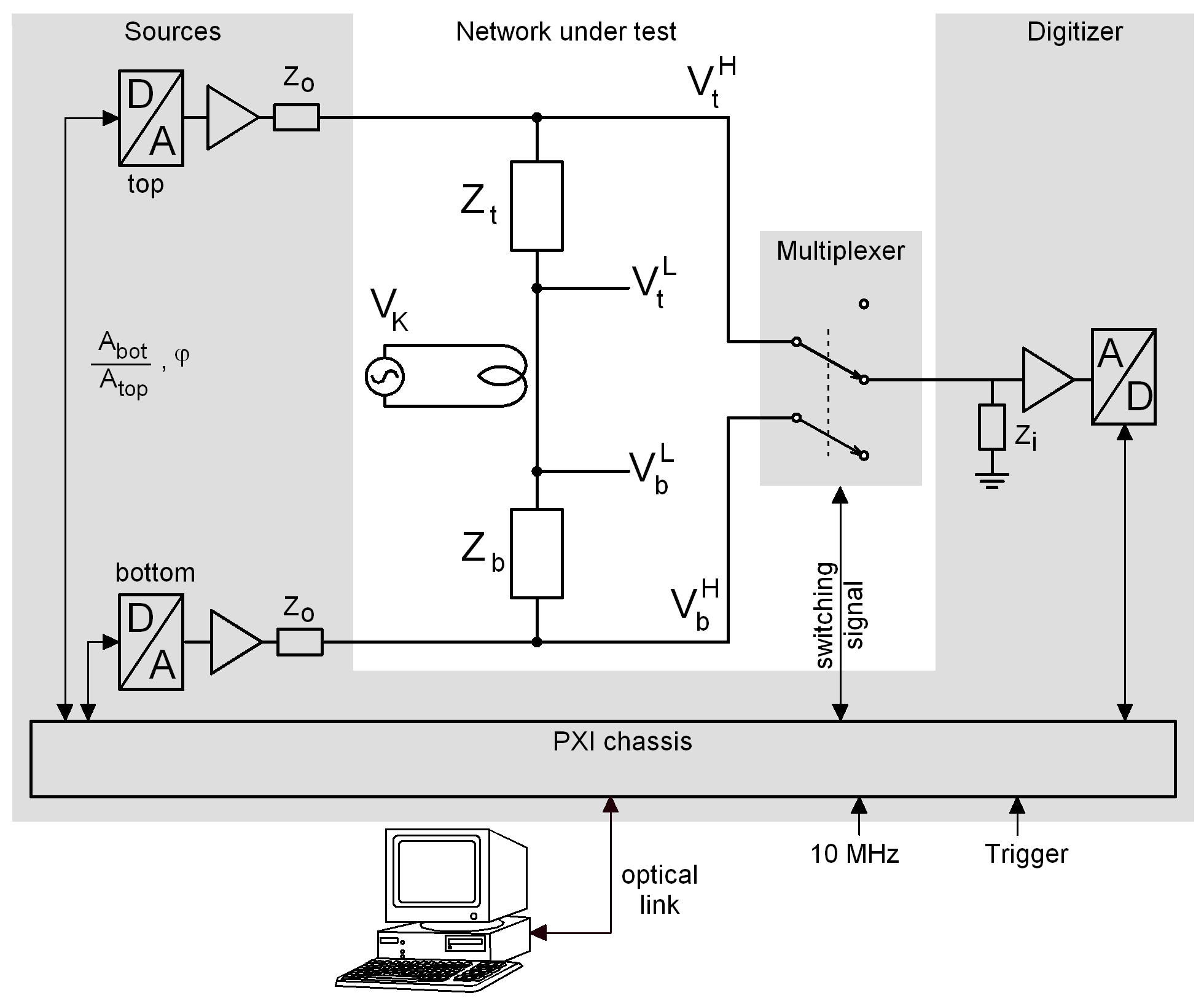}
        \caption{Schematic of the sampling system in a configuration used for the comparison of two impedance standards $Z_{t}$ and $Z_{b}$.}
        \label{Fig_Principle_detail}
    \end{center}
\end{figure*}

In figure \ref{Fig_Principle_detail}, the set-up of the sampling system is shown in
the configuration used for the comparison of a resistance standard $Z_t$ to an
inductance standard $Z_b$. The outer conductor of the coaxial cables has been
omitted for clarity.

The synchronous sampling system is built around a PXI\footnote{PCI  eXtensions for Instrumentation (PXI) is a modular instrumentation platform. See http://www.pxisa.org for more information} chassis having different
modules:
\begin{itemize}
\item The core of the system consists of two high-performance, high-accuracy analog I/O devices
(NI~PXI~4461). Each  device has two digital to analog converters (DAC)  and two analog to digital converters (ADC). Each channel has its own sigma-delta converter with 24-bits of resolution and a maximum sampling rate of 204.8~kS/s. A total of four sinusoidal signals can be synchronously generated with given amplitude and phases relations. In addition, four different signals can be simultaneously measured. 

\item A home-made two channel coaxial multiplexer switches both the 
inner and outer conductors of each channel. This is equivalent to physically connect the digitizer to the different measurement
positions in the network under test. The switching signals are supplied from a relay driver
device (NI~PXI~2567).

\item All devices in the PXI chassis use the same 10~MHz back-plane clock signal which is in turn synchronized to an external 10~MHz reference clock. This guarantees the perfect synchronization of the signal generation and signal digitalization. Moreover, the frequency accuracy of the generated signals is directly given by the accuracy of the 10~MHz reference clock, in our case an atomic Cs-clock.

\item Finally, the data transfer between the PXI modules and the computer is made through an optical link which ensures a good electrical insulation and prevents ground loops. 
\end{itemize}

\section{Balance equation}
Referring to figure \ref{Fig_Principle_detail}, the top and bottom sources supply
sinusoidal voltages to $Z_t$ and $Z_b$ respectively. The amplitude ratio $A_{bot} /
A_{top}$ as well as the phase shift $\varphi$ between the two sources are adjusted (Wagner balance)
to null the voltage, $V_t^L$, at the low voltage port of the top impedance.  A small voltage, $V_K$, is then injected in the link between the low current port of the standards to null the voltage, $V_b^L$, at the low voltage port of the bottom impedance (Kelvin balance). The realization of these two balances ensures
the four terminal-pair definition of the standards \cite{Cutkosky64} and the
ratio of the impedances is then given by the ratio of the measured voltage at the high voltage port of the standards:
\begin{equation}\label{BasicEquation}
      \frac{Z_b}{Z_t}=-\frac{V_b^H}{V_t^H}=-(A+j B)
\end{equation}
Assuming that the bottom impedance is an inductance given by $Z_b=R_s+j \omega L_s$ and the reference top impedance is a resistance given by $Z_t=R_t (1+j \omega \tau)$, it is straightforward to express the series inductance $L_s$
and series resistance $R_s$ in terms of the reference
resistance $R_t$, its time constant $\tau$ and the parameters A and B:
\begin{eqnarray}
    \label{Eq_BasicLsRs}
    L_s=-R_t (\frac{B}{\omega}+A \tau)\\
    R_s=-R_t (A-\omega B \tau)
\end{eqnarray}
$\omega$ being the angular frequency of the measured signals. The dc value of
the reference resistor is measured in terms of the conventional value of the
von Klitzing constant $R_{\rm K-90}$ 
 while its frequency dependence and time constant are evaluated by
direct comparison to a calculable resistance \cite{Gibbings63, Hutzli04}.
Therefore, this calibration chain leads to a realization of the henry in
terms of the ohm and the second. 

The bridge described above gives the four terminal-pair inductance and
resistance of the standard under test while most of the inductance standards on
the market are still defined as two terminal standards. Therefore, a four
terminal-pair (4TP) to two terminal (2T) adapter has to be used.


In addition to the comparison of an inductance to a resistance standard as described above, the sampling system is in fact
able to compare any kind of impedances (R-C, L-C, L-L, R-R or C-C). Theses
measurement capabilities open up a large field of application as well as many
possibilities of accuracy checks through different triangulation schemes.

\begin{figure}[b!]
    \begin{center}
        \includegraphics[width=\mywdithgraph] {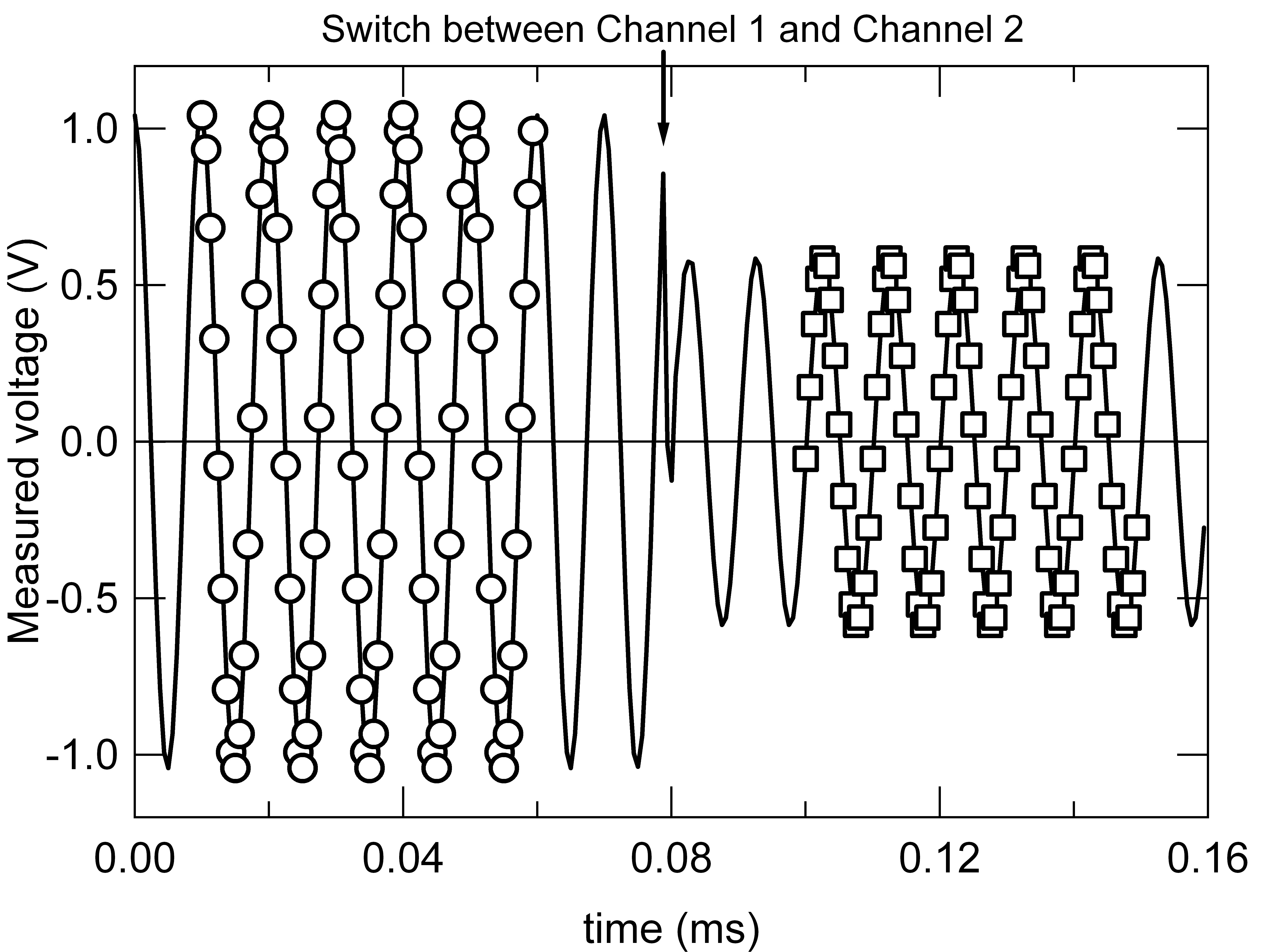}
        \caption{Typical measurement sequence for $m$=8 and $k$=2. Solid line: signal applied to the digitizer.
        Solid circles: the $m$-3 sampled periods of channel 1; solid squares: the $m$-3 periods of channel 2}
        \label{Fig_MeasSequence}
    \end{center}
\end{figure}

\section{Measuement sequence}

A measurement sequence (consisting of two sets of data) is represented in
figure \ref{Fig_MeasSequence}. It is similar to the sequence presented in
\cite{Ramm99} and can be described as follows: The digitizer is acquiring data
at the sampling frequency $f_s=n \cdot f$, where $f$ is the frequency of the
measured signal. A total of 2$m$ periods of the measured signal is sampled in a
sequence, half for channel~1 and half for channel~2.

In order to avoid settling problems due to the switching between the two
channels, $k$ periods before and after the switching are discarded. Of course,
$k$ depends on the switching time of the multiplexer's relays and on the
frequency $f$.

In addition, the first and the last period of the sequence are also discarded.
Since the digitizer is continuously measuring during the sequence, the phase
relation between the two sets of samples is well defined.

Finally, the discrete Fourier transform (DFT) of each data set ($V_t^H$ and
$V_b^H$) is computed and the ratio of the complex Fourier coefficients of the
fundamental is calculated:
\begin{equation}\label{Eq_DFT_ratio}
      \frac{{\rm DFT} \left[ V_b^H \right]_f}{{\rm DFT} \left[ V_t^H \right]_f}=A+j B=r \cdot e^{j\phi}
\end{equation}
The resulting $A$ and $B$ (or $r$ and $\phi$) parameters contain information on both the amplitude
ratio and the phase relation between the fundamental components of the measured
voltages.

Simultaneously to the above sequence, two other detectors (not shown in figure \ref{Fig_Principle_detail}) are monitoring the voltage level at the low voltage port of the impedance standards. The small residual voltages $V_t^L$ and $V_b^L$ are then used to correct, if necessary, the auxiliary balances between two consecutive measurement sequences.

\section{The uncertainty budget}

According to the model equation (\ref{Eq_BasicLsRs}), the relative uncertainty on the measured series inductance $L_s$ is given by

\begin{equation}\label{Eq_uLs}
      \frac{u(L_s)}{L_s} =\Bigg( \bigg[ \frac{u(R_t)}{R_t} \bigg]^2 + \bigg[ \frac{R_s}{L_s} u(\tau) \bigg]^2 + \bigg[ \frac{u(\omega)}{\omega} \bigg]^2+ \bigg[ \frac{R_t}{\omega L_s} u(B) \bigg]^2+ \bigg[ \frac{R_s \tau}{L_s} u(A) \bigg]^2 \Bigg)^{0.5}
\end{equation}
where we used $A=-R_s/R_t$ and $B=-\omega L_s/R_t$ for simplification.

In addition to the five terms of (\ref{Eq_uLs}), supplementary uncertainty components have to be added to take into account for the digitizer linearity, errors in the Wagner and Kelvin balances, current redistribution occurring after the multiplexer switching, connecting cables, errors due to the 4TP-2T adapter and for the measurement of the short ($L_{short}$).

\begin{table*}[h]
\begin{center}
\caption{Combined relative uncertainty (in $\mu$H/H) obtained for the calibration of inductance at frequencies between 50~Hz and 20~kHz. The higher frequency limit for the calibration of 1~H and 10~H standard is given by the resonant frequency $f_r=(LC)^{-0.5}$. The impedance of the 1~$\mu$H standard at frequencies smaller than 1~kHz is so small than the relative uncertainty becomes higher than 10\%. The minimum uncertainties as well as uncertainties smaller than 50~$\mu$H/H are highlighted.}
\begin{tabular}{c|c c c c c c c c}\label{UBudget_table}
$L_{Nominal}$ & 50~Hz  & 100~Hz & 400~Hz & 1~kHz & 5~kHz & 10~kHz & 15~kHz & 20~kHz \\ 
\hline 
1~$\mu$H    & -      &  -     & -       & 17000 & 3500  & 1800  & 1200  & 900  \\
10~$\mu$H   & 33000  &  17000 & 4300    & 1800  & 350   & 180   & 120   & 100   \\
100~$\mu$H  & 3300   &  1700  & 430     & 180   & \lG 35    & \lG 20    & \lG 23    & \lG 35   \\
1~mH        & 330    &  170   & \lG 44      & \lG 18    & \lG 35    & \lG 20    & \lG 23    & \lG 35   \\
10~mH       & \lG 33     &  \lG 18    & \lG 28      & \G 12    & \G 12    & \lG 16    & \lG 25    & \lG 40   \\
100~mH      & \lG 22     &  \lG 18    & \G 12      & \G 12    & \lG 26    & \lG 50   & 120   & 220   \\
1~H         & \lG 22     &  \lG 18    & \G 12      & \lG 13    & 130   & 800  & - & -     \\
10~H        & \lG 22     &  \lG 18    & \G 12      & 60   & - & -     & -     & -     \\
\end{tabular}
\end{center}
\end{table*}

All these uncertainty components are discussed in details in Annex A. Table \ref{UBudget_table} shows the combined relative uncertainties (k=1) obtained for the calibration of standard inductances from 1~$\mu$H to 10~H at frequencies from 50~Hz to 20~kHz.
The smallest uncertainty of about 12~$\mu$H/H is reached at frequencies between 400~Hz and 5~kHz for inductances having an impedance of about 1 to 10~k$\Omega$ at these frequencies. Under such conditions, the main uncertainty component comes from the $1/f$ noise of the sources (DACs) which limits the type A uncertainty on the measured voltage ratio $r$. 

For large inductances at high frequencies, the limiting factor is the effect of the 4TP-2T adaptor on the self-capacitance of the inductance. This effect is not related to the sampling system and can be improved only by physically modifying the terminal configuration of the standard from two terminal to four terminal-pair.

For small inductances at low frequencies, the limiting factor is related to the small voltage ratio $r$ that has to be measured. Indeed, the smallest reference resistor is 10~$\Omega$ which is already more than 100 times larger than the impedance of 100~$\mu$H at 50~Hz. On the other hand, the limited current that can be supplied by the sources ($<$ 5~mA) is also limiting the magnitude of the measured voltage when small impedances are compared. 

Finally, it should be noted that these uncertainties do not include the uncertainty due to the device under test. Most of the time, this uncertainty is dominated by the large temperature coefficient of commercial inductances which can be as large as 30~$\mu$H/H per kelvin. Therefore, the uncertainty level of our sampling system is perfectly suited for calibration of customer's inductances. 

\section{Validation of the measuring system}
The functionality of the new measuring system has been assessed through the inter-comparison EURAMET.EM-S26 based on the measurements of a 100~mH inductance standard at a frequency of 1~kHz. The results obtained during this comparison are in excellent agreement with the results of the other participating laboratories.

\begin{figure}[bth]
    \begin{center}
        \includegraphics[width=\mywdithgraph] {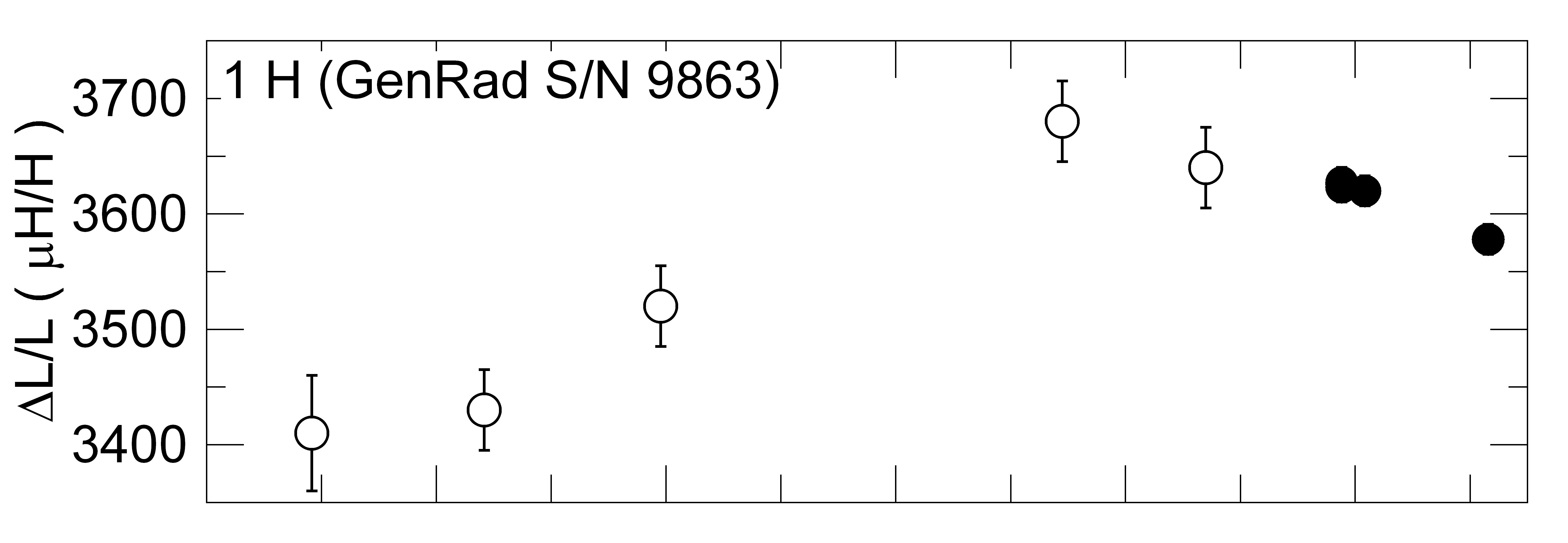}
        \includegraphics[width=\mywdithgraph] {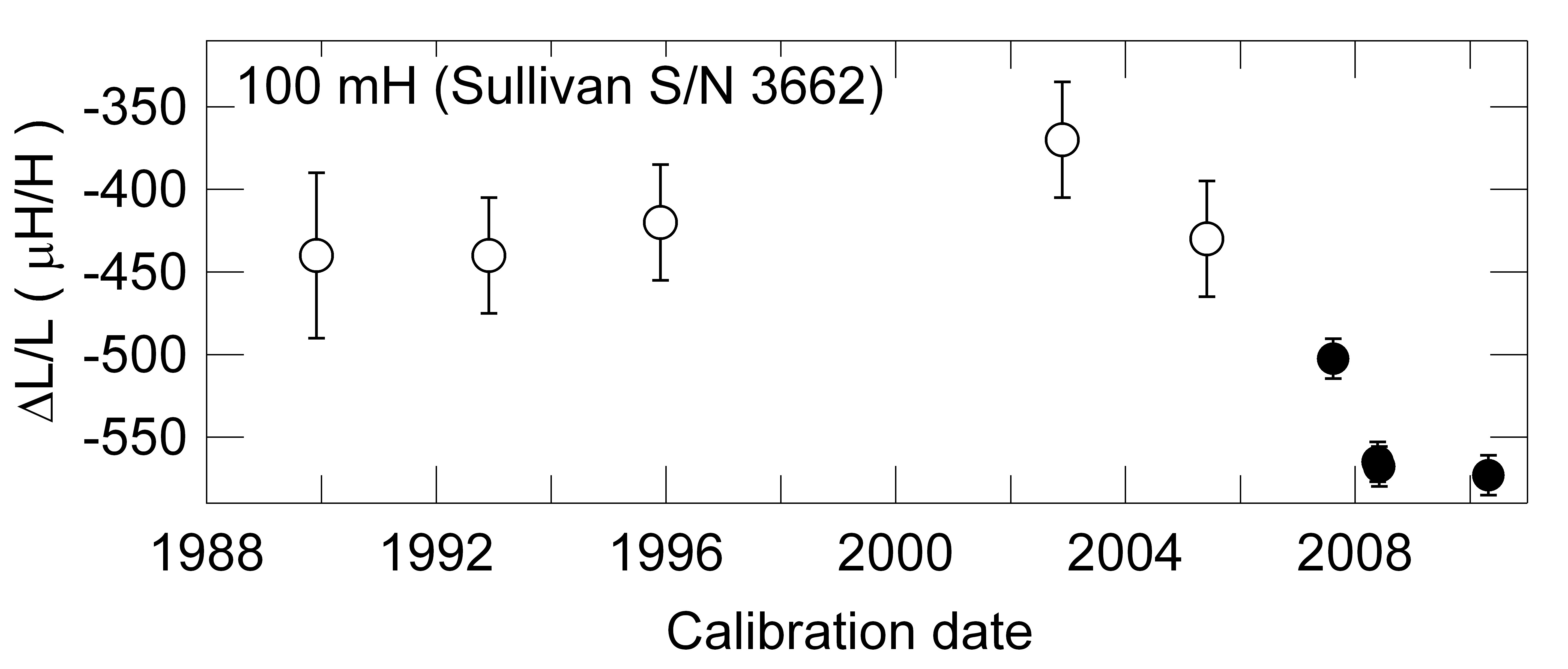}
        \caption{Time evolution of the relative difference of the inductance from its nominal value. Top: for a 1~H standard (GenRad 1482). Bottom: for a 100~mH standard (Sullivan-Griffiths R1940). The open circles are the results of calibrations carried out at NPL while the solid circles are the results of calibrations carried out with the sampling system.}
        \label{Fig_History}
    \end{center}
\end{figure}

Before 2005, the traceability of the inductance calibrations carried out at METAS was realized through a set of inductance reference standards regularly sent to NPL for calibration. Figure \ref{Fig_History} shows the evolution of the relative deviation of a 1~H and 100~mH inductors from their nominal values over a period of more than ten years. For both standards, the results of the calibrations carried out with the sampling system are in agreement with the expected evolution extrapolated from NPL's calibrations. The size of the uncertainty bars in the graph shows the accuracy improvement brought by the new sampling system in the realization of the inductance scale at METAS. 

Moreover, it is of interest to note that the drift of the inductance value dramatically changed around 2003. The inductances were indeed increasing before this time while they start to decrease afterwards. Such a surprising behavior has been observed, with different amplitude, on different inductance standards and is not yet explained. In any case, these instabilities stress the importance of a regular calibration of the reference standards to guarantee the correctness of the inductance measurements.

\begin{figure}[bth]
    \begin{center}
        \includegraphics[width=\mywdithgraph] {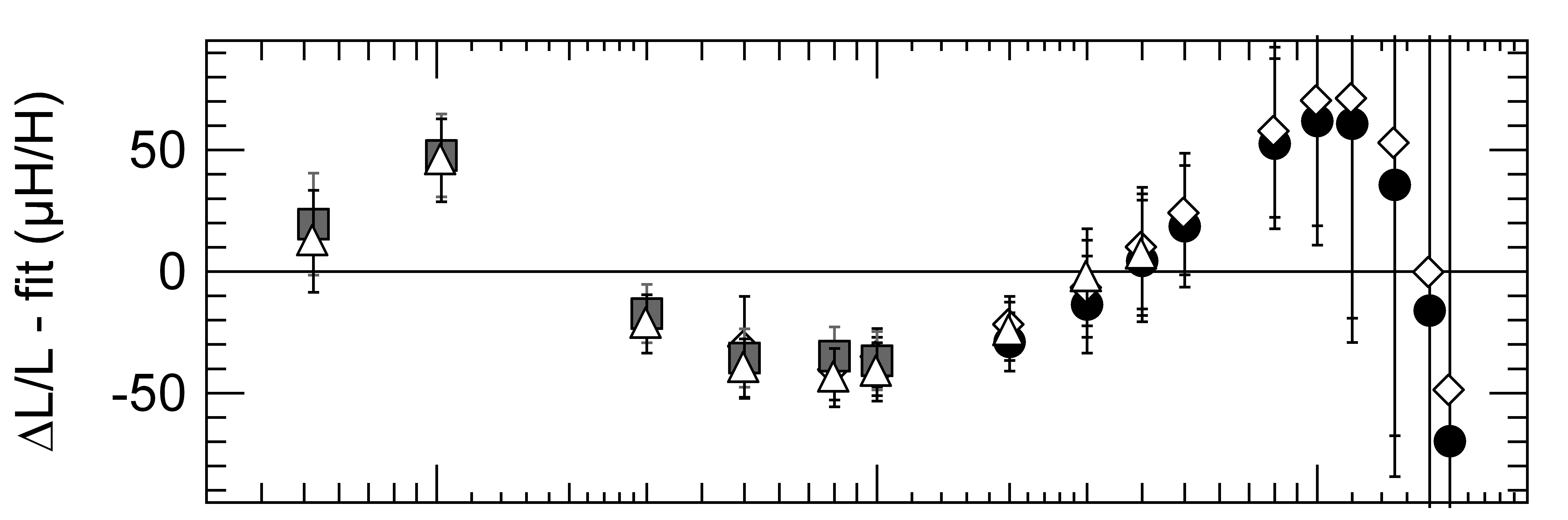}
        \includegraphics[width=\mywdithgraph] {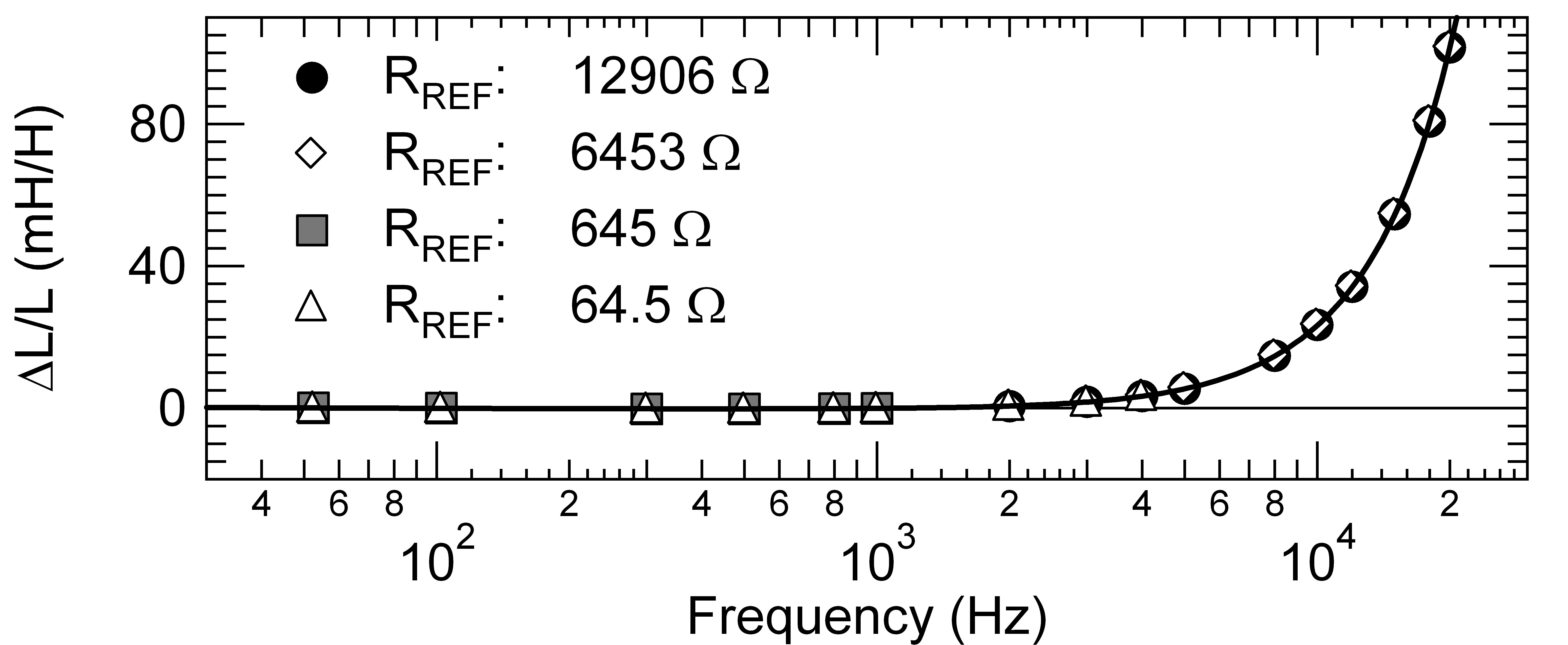}
        \caption{Bottom: Frequency dependence of a 100~mH standard measured using the sampling system at frequencies between 50~Hz and 20~kHz. Depending on the frequency, different resistance standards are used as reference. Top: Difference between the measured value and the least square fit (see text for details).}
        \label{Fig_100mH_FreqDep}
    \end{center}
\end{figure}

A further test of the system has been carried out by measuring a 100~mH inductor at different frequencies between 50~Hz and 20~kHz. Figure \ref{Fig_100mH_FreqDep} shows the frequency dependence of the relative difference of the inductance from its nominal value. The impedance of the inductance increases linearly with the frequency. Therefore, three different resistors have been used as references to maintain the voltage ratio $r$ between 0.1 and 10 over the whole frequency range.

At the highest frequencies, the inductance value increases due to the internal capacitor $C$ that forms a resonant circuit with $L$ (see equation \ref{Eq_Lequivalent}). At the lowest frequencies, eddy currents make the measured inductance to slowly increase. This effect can be accounted for by replacing $L$ by $L_0+L_1/\sqrt{\omega}$ in (\ref{Eq_Lequivalent}), as explained in \cite{Hanke91}. 

The measured frequency dependence has been fitted using the function given by (\ref{Eq_Lequivalent}) with the three parameters $L_0$, $L_1$ and $C$. The resistance $R$ was neglected in this case. The top of figure \ref{Fig_100mH_FreqDep} shows the difference between the measured value and the least square fit. For all frequencies, the inductance value has been measured using different reference resistors. It is important to note that, at any given frequency, the value of the inductance is independent of the reference resistor used as shown in figure \ref{Fig_100mH_FreqDep}. This demonstrates that the measuring system is operational, within the stated uncertainty, over its whole frequency range.

\section{Conclusion}
A new synchronous sampling system has been designed for the calibration of inductances in terms of resistance over the frequency range from 50~Hz to 20~kHz. The use of different reference resistors allows the measurement of inductances from 1~$\mu$H to 10~H. The system is based on commercial PXI boards hosting 24-bits, 204.8~kS/s ADCs and DCAs. The whole system is computer controlled making the automatization of the balancing procedure possible. 

The different sources of uncertainty have been identified and the uncertainty components evaluated in details. The smallest combined uncertainty of about 12~$\mu$H/H ($k$=1) is reached at frequencies between 400~Hz and 5~kHz. The limiting factor is the $1/f$ noise of the sources (DACs) which prevents a perfect Wagner balance and limits the precision on the measurement of the voltage ratio $r$.
For small inductances at low frequencies, the limiting factor is related to the small voltage ratio $r<0.01$ that has to be measured. On the other hand, the limited current that can be supplied by the sources ($<$ 5~mA) is also limiting the magnitude of the measured voltage when small impedances are compared. From this point of view, a better precision could be obtained using buffer amplifiers, with a larger current compliance, at the output of the DACs.

The sampling system presented here is perfectly adapted for routine calibration of inductance standards. Its level of uncertainty is similar to the uncertainty component due to the large temperature coefficient (30~$\mu$H/H per kelvin) of commercially available inductance standards when they are not temperature stabilized.

Moreover, it has been shown that the measured voltage ratio can span the range $0.1<r<10$ without a significant loss of accuracy. Therefore, only a limited number of reference resistors is needed to cover the whole range of inductance (1~$\mu$H to 10~H) over a broad frequency range (50~Hz to 20~kHz). 

An additional advantage of this sampling system is its versatility. Indeed, the kind of impedance to be compared is not limited to inductors and resistors. Any kind of impedances could be compared and the same system is therefore adapted for R-R, R-C, L-L or C-C comparisons, giving this sampling system a great potential of use in many laboratories around the world.

\section*{Acknowledgments}
The authors wish to acknowledge Heinz~B\"{a}rtschi for his technical skills and B.~Jeckelmann for his
continuous support.

\appendix
\section{System characterization and details of the uncertainty budget}

This section present a systematic characterization of the sampling system as well as a detailed discussion of its twelve uncertainty components which are summarized in table \ref{U_table}. As an example, table \ref{U_table} also gives the numerical values ($k$=1) for the case of a 100~mH inductance standards measured at 1~kHz. The type of uncertainty as well as the kind of distribution are also given.

\begin{table*}[h]
\begin{center}
\label{U_table}
\caption{List of the relative uncertainty components in $\mu$H/H when the frequency, $f_{kHz}$, in given in kHz. The symbols are defined in the text. The last column gives the numerical values (k=1) for an inductance of 100~mH measured at 1~kHz.}
\begin{tabular}{c l c l >{$\displaystyle}c<{$} c}
\# & Uncertainty component  & Type & Distribution & \textrm{Expression} & 100~mH at 1~kHz\\ 
\hline \hline
1 & Reference resistor          & B     & Rectangular    & \frac{u(R_t)}{R_t}                                        & 1.2\\
2 & Reference time constant     & B     & Rectangular    & \frac{R_s}{L_s} \cdot u(\tau)                             & 2.0\\
3 & Measuring frequency         & B     & Rectangular    & \frac{u(\omega)}{\omega}                                  & $<$ 0.1\\
4 & Measured value A            & A     & Normal         & \frac{R_s}{L_s}\cdot \tau \cdot u(A)                      & $<$ 0.1\\
5 & Measured value B            & A     & Normal         & \frac{R_t}{\omega L_s} \cdot u(B)                         & 10.3\\
6 & Digitizer linearity         & B     & Rectangular    & (2+0.1 \cdot f_{kHz}^2)\cdot 10^{-6}                            & 1.2\\
7 & Wagner balance              & B     & Rectangular    & D_W \cdot (1+\frac{1}{r}+\frac{Z_t}{Z_i})                       & 5.8\\
8 & Kelvin balance              & B     & Rectangular    & \frac{D_K}{r}                                             & 0.6\\
9 & Switching                   & B     & Rectangular    & \frac{D_S}{r}\frac{Z_o}{Z_o+Z_b}                          & $<$ 0.1\\
10 & Connecting cable           & B     & Rectangular    & 0.15 \cdot 10^{-6} f_{kHz}                                & 0.1\\
11 & Adapter 4TP-2T             & B     & Rectangular    & \frac{\omega^2 L_s}{1-\omega^2 L_s C} \cdot u(\Delta C)   & 0.5\\
12 & Offset                     & B     & Rectangular    & \frac{(0.2+0.26/f_{kHz}^2)\cdot 10^{-9}}{L_s}             & $<$ 0.1\\
\end{tabular}
\end{center}
\end{table*}

\subsection{Reference value of $Z_t$}
A set of reference resistors are used for the calibration of inductance
standards from 1~$\mu$H to 10~H at frequencies from 50~Hz to 20~kHz. 
Each resistor is mounted into a metallic box and is defined as a four
terminal-pair standard. The resistor type (Vishay 101) has been chosen for
their low temperature coefficient to insure a weak effect of the temperature
even if the resistors are not temperature stabilized.

The dc value of each resistance is regularly measured in term of $R_{\rm K-90}$.
The highest measured drift is only 1$\cdot 10^{-6}/$year and is taken into
account for the calculation of the reference value.

The frequency dependence as well as the time constant of each resistance has
been evaluated by comparison to calculable resistors \cite{Gibbings63,
Hutzli04}. The accuracy of the calculated frequency dependence of our
calculable resistors has been confirmed by an international comparison
\cite{Bohacek02} as well by a direct calibration using the ac quantum Hall
effect \cite{Overney06b}.

The accuracy of the determination of the reference resistance at the different
frequencies is typically smaller than 2$\cdot 10^{-6}$ at low frequency and quadratically increases to 20$\cdot 10^{-6}$ at 20~kHz. The typical accuracy on the time constant is less than 5~ns.

\subsection{Uncertainty on the test signal's frequency}
The frequency of the test signal is directly involved in the determination of the inductance (see \ref{Eq_BasicLsRs}) and its accuracy has therefore to be determined. 

The timing information for all ADCs and DACs of the NI~PXI~4461 board \cite{NI4461_spec} comes from the common sample clock timebase signal. A direct digital synthesis (DDS) chip produces the sample clock timebase synchronized with the 10~MHz back-plane clock signal of the PXI chassis. This 10~MHz back-plane clock signal can be, in turn, synchronized with an external 10~MHz reference signal.

The relative accuracy of the test signal frequency has been measured to be better than the 20~$\mu$Hz/Hz when the internal back-plane clock is used and better than 0.05~$\mu$Hz/Hz using an external 10~MHz signal.

\subsection{Uncertainty on the measured voltage ratio}
The measured quantities are the two data sets representing the top and bottom voltages $V_t^H$ and $V_b^H$. The two data sets are coherently sampled ($n$ samples per period) during the same measurement sequence. Therefore, the ratio of the $n_{th}$ coefficient of the DFT of each data set gives the magnitude and the phase of the ratio of the fundamental component of $V_t^H$ and $V_t^H$. Here, it is assumed that a possible error in the DFT calculation is negligible in comparison to the other uncertainty components.

Therefore, the uncertainty components related to the measured ratio are either due to the digitizer itself (stability and non-linearity) or to the deviation from the balance conditions. 

\begin{figure}[b!]
    \begin{center}
        \includegraphics[width=\mywdithgraph] {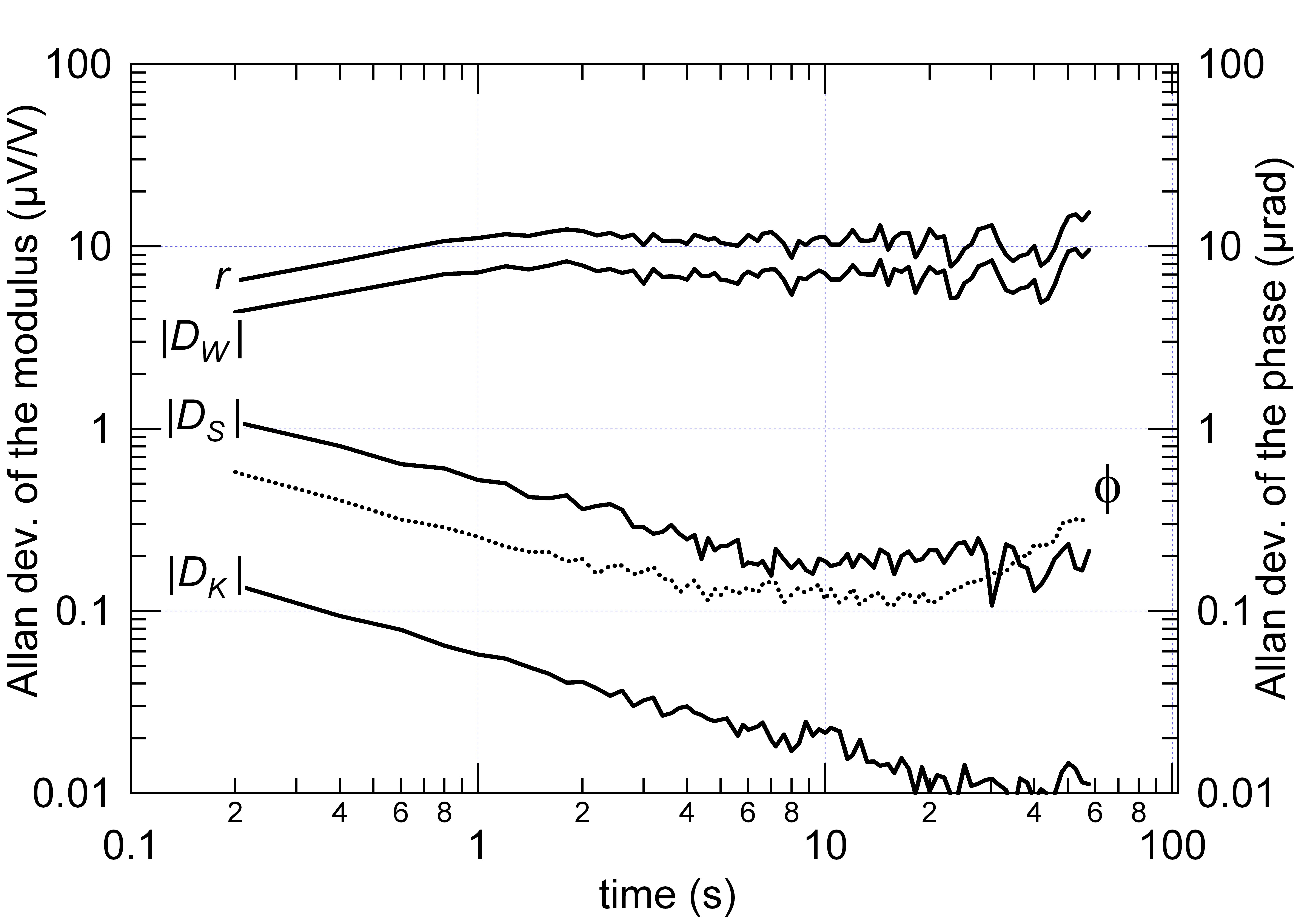}
        \caption{Allan deviation of the voltage ratio (magnitude and phase) measured during the comparison of 100 mH to 645~$\Omega$ at 1~kHz. }
        \label{Fig_UTypeA}
    \end{center}
\end{figure}

To evaluate the Type A uncertainty of the measured voltage ratio ($u(A)$ and $u(B)$), the bridge has been balanced for the comparison of a 100~mH inductor to a 645~$\Omega$ resistor at 1~kHz. The measurement sequence (see figure \ref{Fig_MeasSequence}) has been repeated every 0.2~s during a few minutes. Figure \ref{Fig_UTypeA} shows the Allan deviation of the magnitude $r$ and the phase $\phi$ of the measured voltage ratio.

The constant Allan deviation of $r$ over the whole measurement time $\tau$ indicates that a $1/f$ noise is limiting the uncertainty on the mean value of $r$ to a level of about 10~$\mu$V/V. Such a noise level is more than ten times larger than the noise level of the digitizer itself \cite{Overney10}. The limiting factor is the stability of the voltage sources (output of the DACs of the NI~PXI~4461 board). Regarding the Allan deviation of $\phi$, the $1/f$ noise floor is reached after a measurement time between 5 and 10 seconds. Then the  Allan deviation increases for longer measurement time. Therefore, the lowest uncertainty expected on the mean value of the measured phase is about 0.2~$\mu$rad after 10 seconds of repetition of the measurement sequence.

\subsection{Digitizer linearity}
In the following, it is assumed that the transfer function of the digitizer is given by
\begin{equation}\label{Eq_Vdigitizer}
    V_{out}=V_{in}+a \cdot V_{in}^2+b \cdot V_{in}^3
\end{equation}
where $V_{out}$ is the output voltage of the digitizer $a$ and $b$ are the nonlinearity coefficients. Applying a sinusoidal voltage at the input of the digitizer, $V_{in}=A_{in}\cdot \sin (\omega t)$, the amplitude of the fundamental component of the output voltage is $A_{in} + \frac{3}{4} b A_{in}^3$. Therefore, the ratio of the fundamental components of two sinusoidal signals is

\begin{eqnarray}\label{Eq_Vratio}
    \frac{A_b(1+\frac{3}{4}b A_b^2)}{A_t(1+\frac{3}{4}b A_t^2)} & \approx  \frac{A_b}{A_t}\bigg[1-\frac{3}{4}b (A_t^2-A_b^2)\bigg]\\
      & \approx \frac{A_b}{A_t}\Big[1+\epsilon_{NL}\Big]
\end{eqnarray}
under the assumption that $\frac{3}{4}bA_t^2 \ll1$. It is worthwhile to note that only the coefficient of the cubic term, $b$, appears in the non-linearity error term $\epsilon_{NL}$. Also, the error $\epsilon_{NL}$ is zero when the amplitudes of the measured signals are identical.

The coefficient $b$ can be frequency dependent and is not straightforward to estimate. The maximum integral non-linearity (INL) has been measured using a stepwise triangular signal generated by a programmable Josephson voltage standard (PJVS) \cite{Jeanneret09, Overney10}. For a signal covering the full range of the input scale of the digitizer, a large non-linearity ($<$10~$\mu$V/V) is observed at frequencies smaller than 10~Hz. The INL decreases below 1~$\mu$V/V as the frequency increases to 125~Hz. When the amplitude of the applied signal is only a fraction of the input range, the INL over the whole frequency range tested (1~Hz to 125~Hz) becomes negligible.

\begin{figure}
    \begin{center}
        \includegraphics[width=\mywdithgraph] {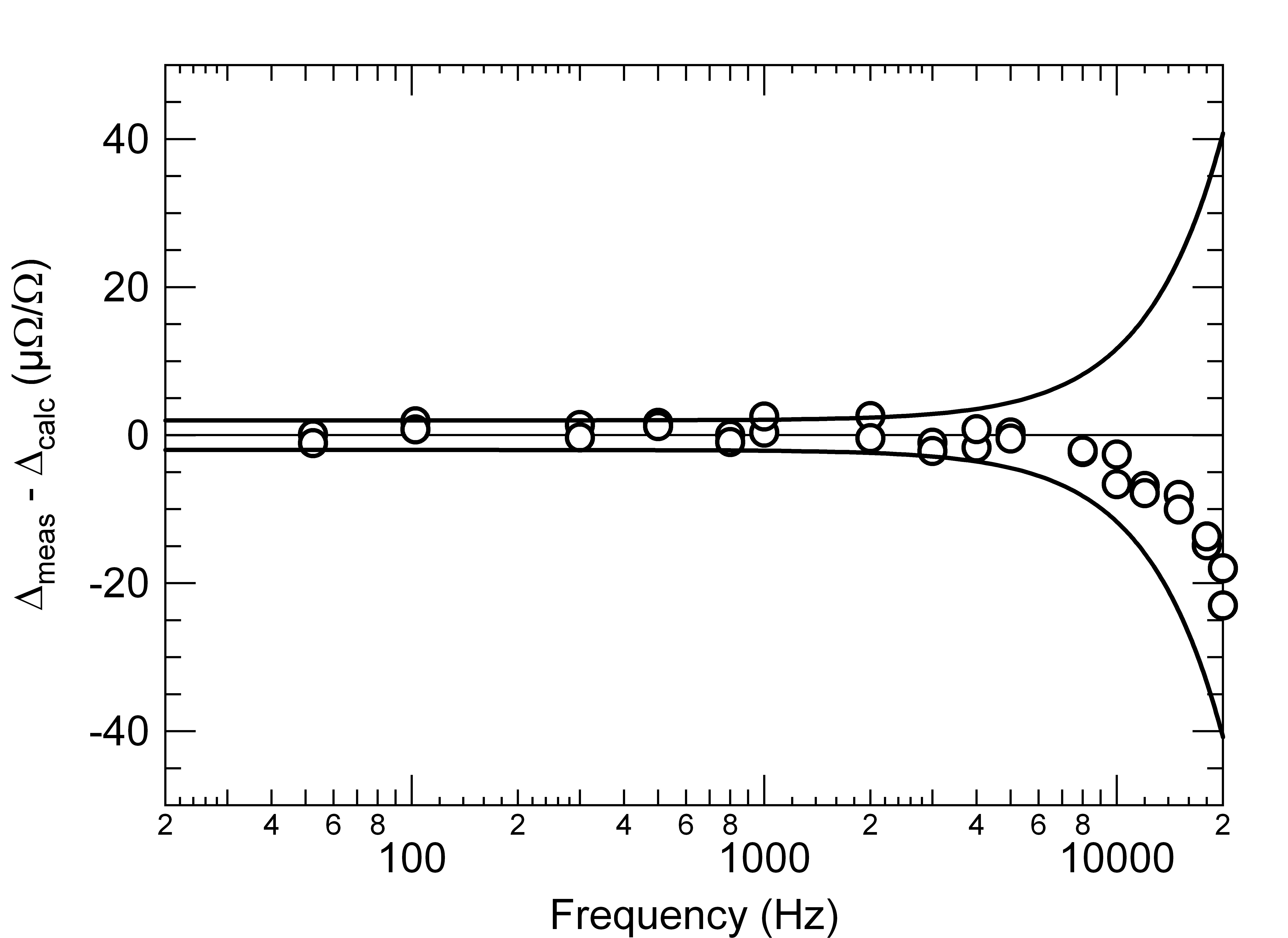}
        \caption{Difference between the measured and the calculated ratio of two calculable resistors. One resistance standard of 1.29~k$\Omega$ and one of 12.9~k$\Omega$. The solid lines represent the uncertainty limits attributed to the error on the measured ratio.}
        \label{Fig_ratioError}
    \end{center}
\end{figure}

To evaluate the effect of the non-linearity on the measured ratio at frequency above 125~Hz, two calculable resistors  \cite{Gibbings63, Hutzli04} of nominal value of 1.29~k$\Omega$ and 12.9~k$\Omega$ have been compared at frequencies from 50~Hz to 20~kHz. Figure \ref{Fig_ratioError} shows the difference between the measured and the calculated resistance ratio (ratio 1/10). The difference is smaller than 2$\mu\Omega/\Omega$ for frequencies below a few kHz and quadratically increases for higher frequencies.
The same measurement has been repeated using calculable resistors of the same nominal value (either two resistors of 12.9~k$\Omega$ or two resistors of 1.29~k$\Omega$). As expected, the error between the measured ratio and the calculated ratio is smaller than 2$\mu\Omega/\Omega$ over the whole frequency range.

The solid line in figure \ref{Fig_ratioError} represents the uncertainty component $u(\epsilon_{NL})$ relative to the non-linearity of the digitizer.

\subsection{Wagner balance}
Referring to figure \ref{Fig_Principle_detail}, the Wagner balance consists in the adjustment of the amplitude ratio and relative phase of the top and bottom sources in such a way that the measured voltage $V_t^L$ is zero. If this balance is not perfectly achieved, the balance equation (\ref{BasicEquation}) becomes
\begin{eqnarray}\label{Eq_WagnerBalance}
      \frac{Z_b}{Z_t}&=-\frac{V_b^H}{V_t^H}+\frac{V_t^L}{V_t^H}\bigg[1+\frac{Z_b}{Z_t}+\frac{Z_b}{Z_i} \bigg]\\
      &= -\frac{V_b^H}{V_t^H}\Big[1+\epsilon_{W}\Big]
\end{eqnarray}
The error in the measured ratio is therefore
\begin{equation}\label{Eq_WagnerError}
    \epsilon_W\leq \bigg| D_W \bigg[1+\frac{1}{r}+\frac{Z_t}{Z_i}\bigg]\bigg|
\end{equation}
where $D_W=V_t^L/V_t^H$ is the residual Wagner balance and $r=|Z_b/Z_t|=|V_b^H/V_t^H|$ is the module of the voltage ratio. In case the Wagner balance is zero, the error $\epsilon_W$ becomes zero and the balance equation (\ref{BasicEquation}) is restored.
The two voltages $V_t^L$ and $V_b^L$ are also synchronously sampled during each measurement sequence using the two ADCs of the second  NI~PXI~4461 board. Therefore, the fundamental component of the relative Wagner signal is also calculated at each measurement sequence. Due to the $1/f$ noise of the sources, the standard deviation on $D_W$ is limited to about 5-7~$\mu$V/V (see figure \ref{Fig_UTypeA}) and will significantly contribute to the uncertainty budget when the ratio of the impedances, $r$, is small (i.e. low inductances at low frequencies).

\subsection{Kelvin balance}
Let us now assume that the Wagner balance is zero. The nominal current $i=V_t^H/Z_t$ flowing through the lead between the low current port of $Z_t$ and the low current port of $Z_b$ will generate a small voltage drop making the Kelvin balance $D_K=(V_t^L-V_b^L)/V_t^H$ different from zero. In this case, the balance equation becomes
\begin{equation}\label{Eq_KelvinBalance}
    -\frac{V_b}{V_t}=\frac{Z_b+Z_K}{Z_t}=\frac{Z_b}{Z_t}\bigg[1+\frac{Z_K}{Z_b}\bigg]
\end{equation}
where $Z_K=(V_t^L-V_b^L)/i=D_K \cdot Z_t$ is the apparent impedance between the low current ports. Therefore, the Kelvin error $\epsilon_K=Z_K/Z_b$ is given by
\begin{equation}\label{Eq_KelvinError}
    \epsilon_K\leq \bigg| \frac{D_K}{r}\bigg |
\end{equation}
A small voltage signal, $V_K$, is injected into the lead between the low current ports of the impedances to balance $D_K$. Figure \ref{Fig_UTypeA} shows that the Allan deviation of $D_K$ is significantly smaller than the noise on $r$ and $D_W$. Since it is correlated, the $1/f$ noise in $V_t^L$ and $V_b^L$ cancel out when the difference is calculated. Although a conservative uncertainty level of 1~$\mu$V/V is attributed to $D_K$, its contribution to the uncertainty budget remains negligible in comparison to the other uncertainty components.

\subsection{Switching effect}
During the Wagner and Kelvin balances, the digitizer is connected to the high potential port of $Z_t$. Therefore, during the first phase of the measurement sequence, the measurement of the voltage $V_t^H$ is carried out in the balance condition. In the middle of the measurement sequence, the switch commutes and the digitizer is then connected to the high potential port of $Z_b$. Due to the finite input impedance of the digitizer, $Z_i$, and the output impedance of the sources, $Z_o$, the current distribution in the whole network is rearranged and the bridge could be slightly out of balance. Therefore, the measured voltage $V_b^H$ could differ from the expected voltage. To avoid this current rearrangement, the channel of the multiplexer, which is not in use, is connected to ground through an impedance equal to $Z_i$. In this way, the two inputs of the multiplexer, which are permanently connected to the network under test, provide a constant load, independent of the channel in use.

Finally, any residual variation in the current flowing through the impedance standards will generate a variation of the measured voltage $V_t^L$ between the two phases of the sequence. It can be shown that the error in the measured voltage ratio is smaller than
\begin{equation}\label{Eq_SwitchError}
    \epsilon_S\leq \bigg| \frac{D_S}{r}\frac{Z_o}{Z_o+Z_b}\bigg |
\end{equation}
where $D_S$ is the difference of the Wagner balance $D_W$ after and before the commutation of the multiplexer.

\subsection{Connecting cables}
The impedance ratio given by equation (\ref{BasicEquation}) is the ratio of impedances including the connecting cables while the quantity of interest is the ratio of impedance as defined at the connector level of both standards. A small correction has, therefore, to be applied to the measured ratio to take into account these connecting cables. The correction can easily be calculated using the lumped parameters of the cables \cite{Kibble84}. However, the correction on the ratio of impedances is close to be the difference of the correction of the impedances alone. Therefore, choosing cables of the same length to connect $Z_t$ and $Z_b$ to the measuring system leads to a negligible correction on the ratio $Z_b/Z_b$. In our setup, the maximum uncertainty on the measured ratio $r$ related to the connecting cables increases linearly with the frequency and amounts to 3~$\mu \Omega/\Omega$ at 20 kHz.

\subsection{4TP-2TP adapter}
\begin{figure}[bth]
    \begin{center}
        \includegraphics[width=0.8\mywdithgraph] {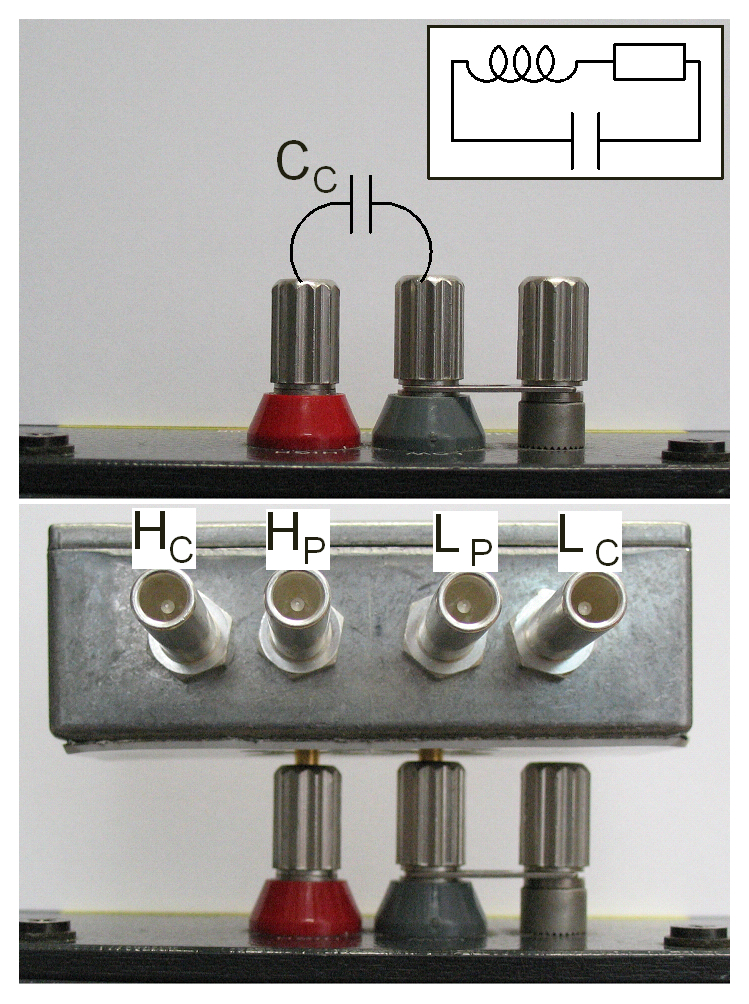}
        \caption{Top: Typical configuration of the terminals of an inductance standard. The capacitance $C_C$ between terminal is part of the standard. The equivalent circuit of the standard is shown in the inset. Bottom: The adapter needed to measure the 2~terminal standard with a for terminal-pair bridge. Any variation of the terminal capacitance $C_C$ will modify the measured value of the inductance.}
        \label{Fig_Adapter}
    \end{center}
\end{figure}

A real inductance standard can be represented by a resistor, $R$, and an inductor, $L$, in series together with a capacitor, $C$, in parallel as shown in the inset of figure \ref{Fig_Adapter}.
The apparent inductance of such a network is given by:
\begin{equation}\label{Eq_Lequivalent}
    L_s=\frac{L(1-\omega^2 L C)-C R^2}{(1-\omega^2 L C)^2+(\omega R C)^2}
\end{equation}
where the capacitance $C$ is in fact the sum of the capacitance between the spires forming the inductance (inside the inductor enclosure) and the capacitance between the connecting terminal $C_C$ (outside the enclosure). Therefore, any variation of $C_C$ due to the adaptor used to connect the measurement cables to the inductor will slightly change the total capacitance $C$ and modify the apparent inductance, $L_s$, according to:
\begin{equation}\label{Eq_DLs}
    \frac{\Delta L_s}{L_s}\approx \frac{\omega^2 L}{(1-\omega^2 L C)}\cdot \Delta C
\end{equation}
Figure \ref{Fig_Adapter} shows the homemade adapter needed to carry out the 4TP measurement of the 2T inductance standard. The adapter has been designed to minimize its influence on $C_C$ which has been measured to be stable within a uncertainty of $u(\Delta C)=$0.2~pF.

\subsection{Offset}

\begin{figure}[bth]
    \begin{center}
        \includegraphics[width=\mywdithgraph] {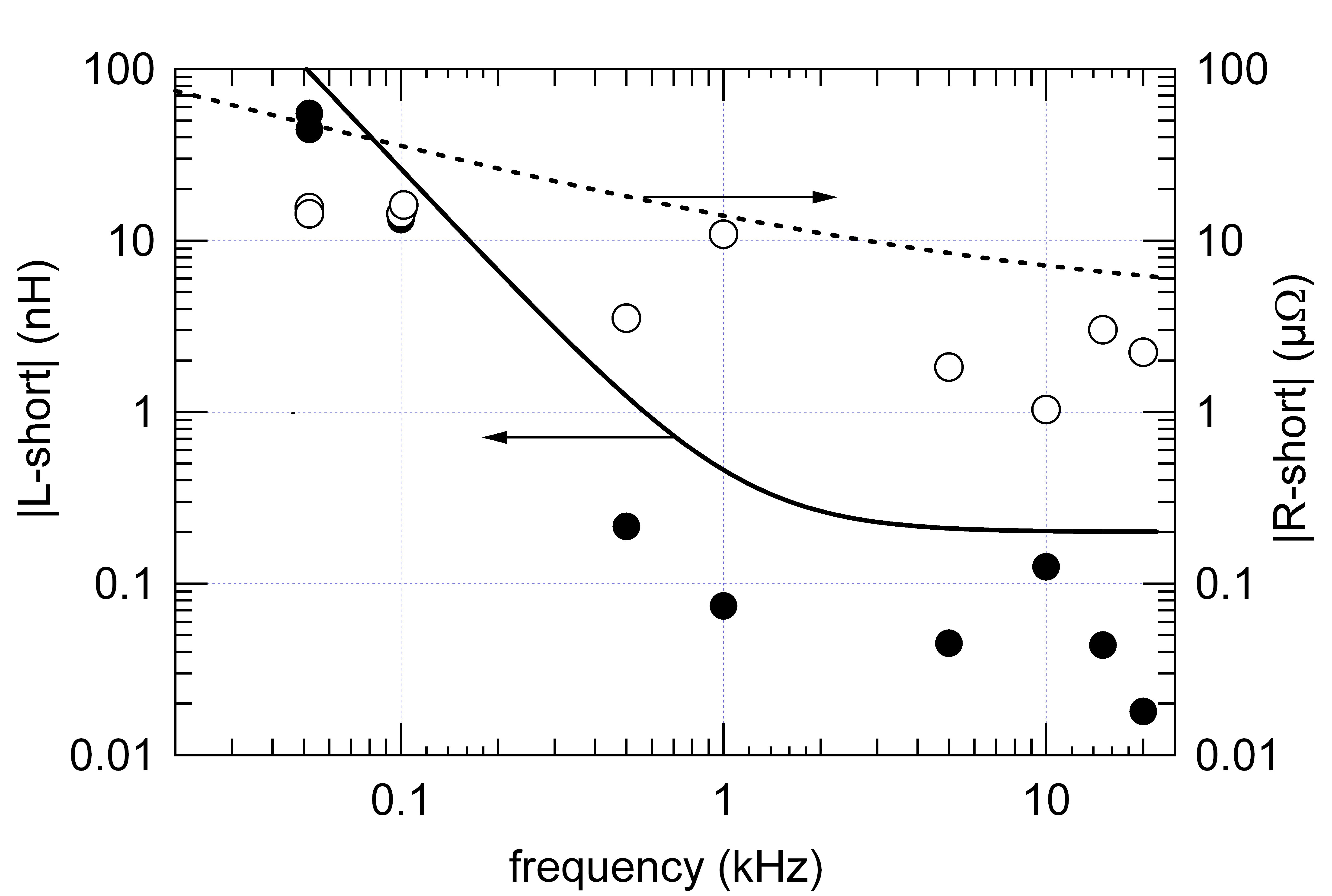}
        \caption{Offset of the measuring system obtained by comparing a zero impedance standard to a 10~$\Omega$ standard at frequencies from 50~Hz to 20~kHz. The solid line and the dashed line represent the maximum offset of the measuring system that has to be taken into account in the uncertainty budget.}
        \label{Fig_Offset}
    \end{center}
\end{figure}

The uncertainty on the measurement of the shorted inductance is of importance especially when low value inductors are measured. To characterize the system for low impedance measurements, a good four terminal-pair zero impedance \cite{Kibble84} has been measured with respect to a 10~$\Omega$ reference standard.
Figure \ref{Fig_Offset} shows the residual inductance and resistance as a function of the frequency. The solid line and the dashed line represent the maximum offset of the measuring system that has to be taken into account in the uncertainty budget.

\section*{References}

\end{document}